\documentclass{pasj00}

\SetRunningHead{Deguchi et al.}{SiO Maser Spectra of  V407 Cyg after the Nova Outburst}
\title{SiO Maser Spectra of  V407 Cyg after the 2010 March Nova Outburst}

\Received{2010/07/14}
\Accepted{2010/11/27 ;PASJ ver 2.9, Nov. 23, 2010}

\author{Shuji \textsc{Deguchi}\altaffilmark{1,2},  Kazutaka \textsc{Koike}\altaffilmark{1,2}, Nario \textsc{Kuno}\altaffilmark{1,2}}
\author{Noriyuki \textsc{Matsunaga}\altaffilmark{3}, Jun-ichi \textsc{Nakashima}\altaffilmark{4}}
\and
\author{Shigeru \textsc{Takahashi}\altaffilmark{1}}
 \altaffiltext{1}{Nobeyama Radio Observatory, National Astronomical Observatory,\\
              Minamimaki, Minamisaku, Nagano 384-1305}    
\altaffiltext{2}{Graduate University for Advanced Studies, 
National Astronomical Observatory, \\
Minamimaki, Minamisaku, Nagano 384-1305}
\altaffiltext{3}{Kiso Observatory, The University of Tokyo, 
 \\ 10762-30 Mitake, Kiso, Nagano 397-0101}
\altaffiltext{4}{Department of Physics, University of Hong Kong, Pokfulam Rd, Hong Kong, China}
 \author{[PASJ Letter 63 No. 1 (Feb. 25, 2011 issue) in press] }
\begin{document}
\maketitle

\begin{abstract}
We report on the time variation of  SiO maser emission  from the symbiotic stellar system
V407 Cyg after the classical nova outburst on 2010 March 10.   
Although both the SiO $J=1$--0 $v=1$ and 2 lines at 43.122 and 42.821 GHz were  found previously
 in the envelope of a mira in the binary system, 
 only weak emission of the $J=1$--0 $v=2$ line has continuously been detectable after the nova outburst.
The line profile exhibited a dramatic change several weeks after the burst;  the component on the higher-velocity side
of the systemic velocity disappeared two weeks after the burst, and a new persistent component appeared on the lower-velocity side later. 
These observations indicate that the SiO emitting regions are wiped out in a time scale of two weeks
by the nova shock, but a part of the masing region is quickly replenished by cool molecular gases expelled by  
the mira pulsation. 
\end{abstract}

\section{Introduction\label{sec:Intro}}
Classical novae are thermonuclear runaway explosion, which occurs in a hydrogen-rich envelope of a white dwarf 
accreting matters from a donor star in a binary system \citep{sta00}.
The donor star is often a main-sequence star or red giant, which fills its Roche lobe. However, 
when  the donor is an mass-losing evolved star detached from the Roche lobe,
the nova shock  interacts with the preexisting cool envelope of the donor and causes various complex phenomena 
(see, e.g., \cite{kwo97,bod06}). 

The symbiotic star system V407 Cyg was found in nova outburst  on 2010
March 10.\footnote{see Central Bureau Electronic Telegrams:
CBET 2199, 2204, and 2205.} 
The donor star in this system is known to be a mira  [an  evolved star
at the asymptotic giant branch (ABG) phase of stellar evolution] with a
period of the optical light curve of $\sim 745$ d  \citep{tat03,shu07}. The long pulsation period  suggests that it is relatively luminous 
compared with typical optical miras in the Galactic disk with a period of about 300--400 d (for example, see \cite{whi00}).
This star is classified to a D-type symbiotic star due to dust-richness in the mira envelope \citep{sea95}.
A summary of optical observations at the time of 2010 March nova outburst  is found in the
American Association of Variable Star Observers ( AAVSO)  web page.\footnote{
http://www.aavso.org/news/v407\_cyg.shtml}
The radio continuum has been detected at various frequencies from 80 GHz down to 1.4 GHz
\citep{poo10,bow10,gir10,gaw10,nes10}. A probable detection of $\gamma$ ray was also reported \citep{che10}.

SiO maser lines have previously been detected in the envelope of the mira in this system \citep{deg05a}.
In general, SiO masers are emitted in a circumstellar envelope at a few stellar  radii ($\sim 10^{14}$ cm)  of the AGB star photosphere.
If we assume the expansion velocity of the nova outburst to be 1000 km s$^{-1}$,
the shock propagates through the SiO masing region in a time scale of 12 days.     
The line profile of  SiO maser emission is expected to vary after the nova outburst in this time scale, 
in which the time variation of the line profile can feasibly be monitored. Therefore,
if we assume the maser emission is terminated due to the passage of a shock front over the masing region,
we can tail  the propagation of the shock wave by monitoring the maser line profile. 
VLBI (very long baseline interferometry) observations may directly image the interaction  
of the shock with the neutral gas in the mira envelope.

In this paper, we present single-dish monitoring of variation of SiO maser line profiles in V407 Cyg
after the nova outburst in 2010 March.
The "classical" nova phenomenon was observed for the first time
in the  SiO maser line in the present research.\footnote{ Note that SiO masers have been detected toward the
M-type supergiant V838 Mon in 2005 \citep{deg05b}, 3 years after a nova-like eruption. 
However, it is believed that the nova-like eruption of V838 Mon was caused
by a  physical mechanism totally different from a classical nova,  e.g., the merger of two stars
(called  "red nova" phenomena; see \cite{sok03,rau07}).} 

\section{Observations and results\label{sec:Obs}}
The monitoring observation of V407 Cyg  was started with the Nobeyama 45-m telescope first on 2010 March 16, in
6 days after the discovery of the nova outburst. Since then, we monitored the object
in the SiO maser lines ($J=1$--0,  $v=1$ and 2) at 43.122 and 42.821 GHz, respectively, once
in every few days in the initial phase, and almost once in every week at the later phase till the end of 2010 May.
The half-power full beam width (HPFBW) of the telescope was about 40$''$ at 43 GHz. 
We used a cooled HEMT receiver (H40) giving $T_{\rm sys}=$ 180--250 K
and acousto-optical spectrometers with high (40 kHz; AOS-H) and
low (250 kHz; AOS-W) resolutions having 2048 channels each.
The spectrometers AOS-H and AOS-W
covered  velocity ranges of $\pm 150$ km s$^{-1}$ and  $\pm 800$ km s$^{-1}$
 with an effective velocity resolution of 0.3 and  1.8 
km s$^{-1}$ per binned channel,  respectively.
The conversion factor of the antenna temperature ($\equiv T_a^*$) to the flux density was
$\sim 2.9$ Jy K$^{-1}$. The typical on-source integration time for these observations was about 10 minutes.
The telescope pointing was checked at the beginning of observations using IRAS 21086+5238, 
a strong nearby SiO maser source. The peak intensities of this SiO maser source are approximately 3.7 and 5.9 K for
the SiO $J=1$--0 $v=1$ and $v=2$ lines, respectively 
(in the later phase, the $v=1$ line intensity increased to 4.5 K, but the $v=2$ intensity stayed almost the same).
Because the SiO maser intensity of the above pointing source was always confirmed
to be nearly constant, we believe that the intensity calibration of the V407 Cyg at 43 GHz using the chopper-wheel method 
was made properly and the uncertainty of the intensity scale is estimated to be approximately $\pm 10$ percent
throughout the period of observations. 
The spectrometer arrays also covered the SiO $J=1$--0 $v=0$ and $v=3$  lines
at 43.424 GHz and 42.519 GHz, respectively, the $^{29}$SiO  $J=1$--0 $v=0$ line at 42.880 GHz,
and H53$\alpha$ at 42.952 GHz. However, none of these lines were detected through all the 
observations in 2010. The rms noise levels for these lines are similar to those of the SiO $J=1$--0,  $v=1$ and 2 lines.
In addition to the 43 GHz observations, we tried to detect the SiO $J=2$--1 $v=1$ and 2 lines
at 86.243 GHz and 85.640 GHz, respectively, and the  $^{29}$SiO  $J=2$--1 $v=0$ line
at 85.759 GHz, with an SIS-mixer receiver (T100) on 2010 March 20 and April 26, 
but none of them were detected.  
We also attempted  to detect the H$_2$O $6_{16}$--$5_{23}$ maser line at 22.235 GHz
on 2010 March 16 and March 20 using the HEMT receiver (H22), resulting no detection.
We summarize the results of these observations in Table 1, and we will not
 discuss negative detections in the rest of this paper. 

\begin{figure*}
\begin{center}
\FigureFile(170mm,100mm){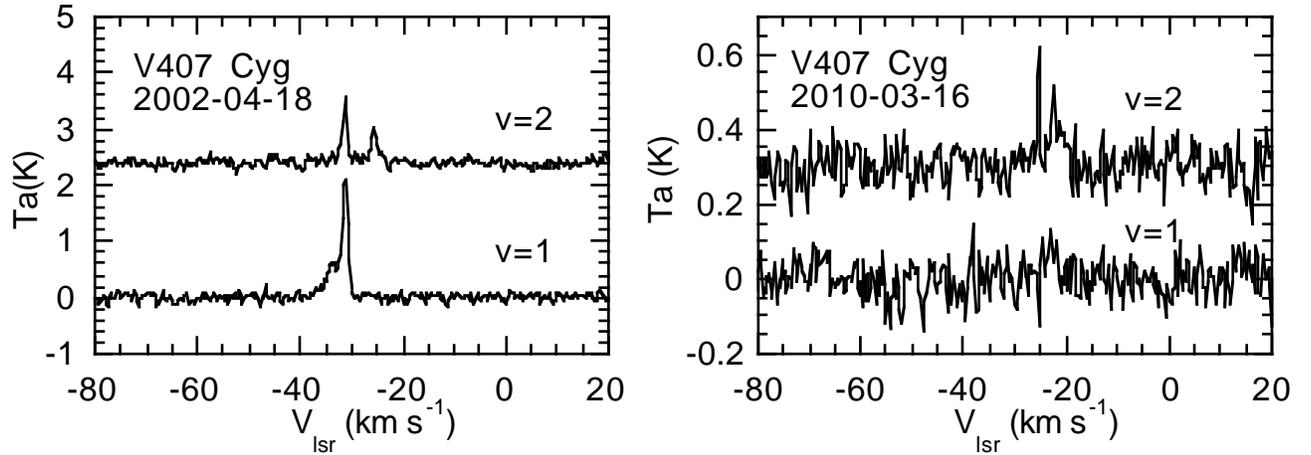}
\end{center}
\caption{Comparison of the spectra of the SiO $J=1$--$0, v=2$ and $v=1$ lines 
on 2002 April 18 (left) and on 2010 March 16 (right).
\label{fig:1}
}
\end{figure*}

\begin{table*}
  \caption{Summary of SiO observations}\label{tab:1}
 \begin{center}
  \begin{tabular}{llccccc}
  \hline\hline
Obs. Date (UT) & Transition & Frequency & $T_a$(peak) &  $V_{\rm LSR}$(peak) & Integ. Intensity & rms \\
(yyyy/mm/dd.d) &            &  (GHz)    &  (K)       &  (km s$^{-1}$)        & (K km s$^{-1}$) & (K) \\
\hline
2010/03/16.0     &  SiO $J=1$--0 $v=1$      &  43.122  & 0.153 & $-$23.1 & 0.220 & 0.045 \\
               &  SiO $J=1$--0 $v=2$      &  42.821  & 0.347 & $-$25.4 & 0.710 & 0.048 \\
               &  H$_2$O $6_{16}$--$5_{23}$  &  22.235  & ---  & --- & --- & 0.021 \\
2010/03/19.0     &  SiO $J=1$--0 $v=1$      &  43.122  & --- & --- & --- & 0.036 \\
               &  SiO $J=1$--0 $v=2$      &  42.821  & 0.186 & $-$22.4 & 0.751 & 0.037  \\
2010/03/20.0     &  SiO $J=2$--1 $v=1$      &  86.243  & ---  & --- & --- & 0.030 \\
               &  H$_2$O $6_{16}$--$5_{23}$  &  22.235  & ---  & --- & --- & 0.018 \\
2010/03/22.0     &  SiO $J=1$--0 $v=1$      &  43.122  & 0.103 & $-$23.0 & 0.402 & 0.029 \\
               &  SiO $J=1$--0 $v=2$      &  42.821  & 0.163 & $-$21.8 & 0.699 & 0.031 \\
2010/03/28.0     &  SiO $J=1$--0 $v=1$      &  43.122  & ---  & --- & --- & 0.037 \\
               &  SiO $J=1$--0 $v=2$      &  42.821  & 0.138 & $-$26.8 & 0.391 &  \\
2010/04/06.0     &  SiO $J=1$--0 $v=1$      &  43.122  &  --- & --- & --- & 0.027 \\
               &  SiO $J=1$--0 $v=2$      &  42.821  & 0.153 & $-$30.0 & 0.441 & 0.027 \\
2010/04/11.0     &  SiO $J=1$--0 $v=1$      &  43.122  & --- & --- & --- & 0.040 \\
               &  SiO $J=1$--0 $v=2$      &  42.821  & 0.196 & $-$30.0 & 0.369 & 0.038 \\
2010/04/13.0     &  SiO $J=1$--0 $v=1$      &  43.122  & ---  & --- & --- & 0.040 \\
               &  SiO $J=1$--0 $v=2$      &  42.821  & 0.251 & $-$30.3 & 0.372 & 0.043 \\
2010/04/19.1     &  SiO $J=1$--0 $v=1$      &  43.122  & ---  & --- & --- & 0.036  \\
               &  SiO $J=1$--0 $v=2$      &  42.821  & 0.183 & $-$29.9 & 0.314 & 0.037  \\
2010/04/26.1     &  SiO $J=2$--1 $v=1$      &  86.243  & ---  & --- & --- & 0.013 \\
2010/04/28.0     &  SiO $J=1$--0 $v=1$      &  43.122  & 0.096 & $-$27.4 & 0.119 & 0.032 \\
               &  SiO $J=1$--0 $v=2$      &  42.821  & 0.176 & $-$30.2 & 0.403 & 0.033 \\
2010/05/05.9     &  SiO $J=1$--0 $v=1$      &  43.122  & --- & --- & --- & 0.036 \\
               &  SiO $J=1$--0 $v=2$      &  42.821  & 0.204 & $-$30.3 & 0.552 & 0.041 \\
2010/05/16.9     &  SiO $J=1$--0 $v=1$      &  43.122  & --- & --- & --- & 0.036   \\
               &  SiO $J=1$--0 $v=2$      &  42.821  & 0.168 & $-$30.0 & 0.440 & 0.035\\
2010/05/25.9     &  SiO $J=1$--0 $v=1$      &  43.122  & ---  & --- & --- & 0.028  \\
               &  SiO $J=1$--0 $v=2$      &  42.821  & 0.188 & $-$29.5 & 0.468 & 0.029  \\
\hline
  \end{tabular}
  \end{center}
\end{table*}


\subsection{Variation of the spectra.\label{sec:velocity}}

Figure 1 shows a comparison of the SiO spectra on 2010 March 16 (right) with those taken
8 years ago (left) using the same telescope \citep{deg05a}. In the previous spectra \citep{deg05a},
the $J=1$--0 $v=1$ line exhibits a single-peak profile  
at  $V_{lsr}=-31.0$ km s$^{-1}$ with an extended tail on the lower velocity side of the peak, and the $J=1$--0 $v=2$ line
exhibits a double-peak profile peaked at $V_{lsr}=-31.3$ km s$^{-1}$ and $-25.9$  km s$^{-1}$.
In contrast, the intensity of the SiO emission observed on 2010 March 16 is clearly weaker than
the previous one, and the line profiles are also different.
The $J=1$--0 $v=1$ line is considerably weaker than the  $J=1$--0 $v=2$ line.\footnote{
Although the line profile generated by the AOS-H spectrometer is less certain in  Figure 1, 
we confirmed the detection of the $J=1$--0 $v=1$ line using 
the AOS-W spectrometer, which has a larger band width (0.25 MHz; $\Delta v\sim 1.7$ km s$^{-1}$).}
Because the profile of the $J=1$--0 $v=1$ line is less certain in the high resolution spectrum of AOS-H, 
we discuss only the variation of the $J=1$--0 $v=2$ line
in  later sections. It is notable that only the higher-velocity components
($V_{lsr}> -26$ km s$^{-1}$) in the $v=2$ line were detectable on March 16.

Figure 2 shows  the time variation of the  $J=1$--0 $v=2$ line profiles over 2.5 months since the nova outburst.  For the
first two weeks (till March 22), the higher-velocity components on the red-shifted side 
had been seen.  The higher-velocity components are progressively weakened, showing a minor change in the profiles as time advances.
 For example,  the sharp peak observed on March 16 at  $V_{lsr}=-25$ km s$^{-1}$ was not seen in the later spectra.
Then, new peaks appeared on the low-velocity side ($V_{lsr}< -26$ km s$^{-1}$)  
after March 22, though these peaks  could not be seen temporarily on April 6 (presumably due to a noise).
After April 11, the low velocity component at  $V_{lsr}=-30$ km s$^{-1}$ maintained nearly 
the same intensity, and also maintained almost the same radial velocity till May 16,
though the peak velocity moved a little to the higher-velocity direction on May 25.
Also the emission which tailed on the higher-velocity side of $V_{lsr}=-30$ km s$^{-1}$ appeared from April 19.
This tailed emission increased in intensity, and then split into two peaks on May 25.    

\begin{figure*}
\begin{center}
\FigureFile(130mm,200mm){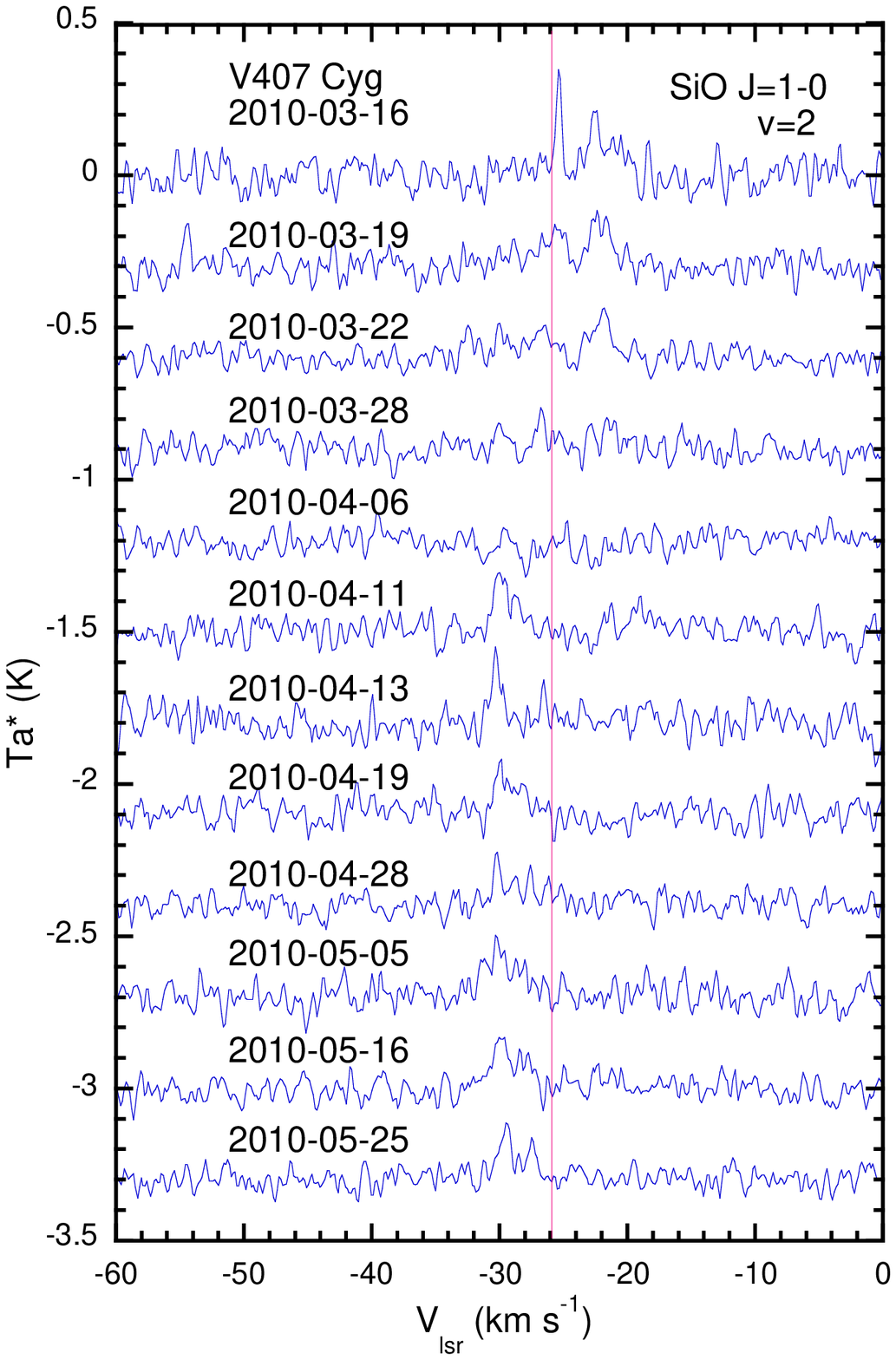}
\end{center}
\caption{Variation of the spectra of the SiO $J=1$--$0  v=2$  line 
from 2010 March 16 to 2010 May 25. The observed dates are shown
on the left of each spectrum in the yymmdd.d format in the Universal Time. 
The vertical red line at $V_{lsr}=-26$ km s$^{-1}$ is  a guide for eye.  
\label{fig:2}
}
\end{figure*}
The intensity-weighted average of the radial  velocity of the SiO $J=1$--0 $v=1$ and 2 lines is usually  a good indicator
of a stellar velocity (e.g., \cite{jew91,jia95}) for evolved stars with SiO maser emission. However,
since V407 Cyg shows a double-peak profile, which is different from a typical SiO maser source,
the intensity-weighted velocity does not seem to be a reliable probe of stellar velocity. 
Alternatively,  we think that the average velocity of the lowest and highest velocities of the emission
(e.g., edges of the line profile at $\sim -32$ and $-19$ km s$^{-1}$)
better represents the stellar velocity, i.e., $V_{\rm LSR}(star)=-26$  km s$^{-1}$,
because the SiO $v=2$ emission is formed in the inner part of the circumstellar envelope which is considerably clumpy. 
However, note that this is the stellar velocity of the mira in the system, because the $v=2$ emission is attached to
the inner circumstellar envelope of the mira. 
This velocity  ($=-42$ km s$^{-1}$ in the Heliocentric radial velocity)\footnote{
This is close to the mira radial velocity of $-41$ km s$^{-1}$, which was given by \citet{tat03b} using the average velocity of the  Ca I $\lambda$ 6573A absorption line
during 1995 and 2001. But this is rather fortuitous if we consider the $\sim 43$ yr orbital motion of the mira of  $\sim 4$ km s$^{-1}$ in the binary model described in the next section.} 
is larger than most of the optically obtained radial velocities of 
atomic lines at the time of burst, which are in the range between $V_{\rm helio}=-50$ and $-60$  km s$^{-1}$
[see Supporting Online Material for \citet{the10}].
 These optical lines (a narrow central part of the profile) are formed in a part of mira wind faced on the white dwarf,
 where the gas is ionized by the UV radiation at radiation front of the shock.
Therefore, the radial velocity of the whole system is likely to be blueshifted 
relative to the mira.

  In addition to the present observations,
 spectra at 43 GHz SiO and 22 GHz H$_2$O masers taken three month before the outburst
(2009 December)	
  became available recently \citep{cho10}. 
  Interestingly, the  SiO $J=1$--0 $v=1$ and 2 spectra  exhibited  the strongest peaks
at $V_{\rm LSR}=-28$  km s$^{-1}$ at that time, but the H$_2$O spectrum at $V_{\rm LSR}=-32$  km s$^{-1}$.
The SiO $J=1$--0 $v=2$ profile showed the other weaker peaks at $ -30$, $-26$, and $-24$ km s$^{-1}$.

These observations before the burst together with our SiO maser observations after the burst reasonably
suggest that the nova outburst terminated the masers progressively from  the lower ($V_{\rm LSR}<-26$  km s$^{-1}$)
to the higher velocities ($V_{\rm LSR}>-26$  km s$^{-1}$).
However,  a month later, a new lower velocity component was created.
Based on these observations, we construct a possible model to explain the variation of the SiO maser profile
in the next section.

\section{Discussion\label{sec:disc}}

The basic physical parameters of the  V407 Cyg symbiotic system  have been derived from the optical/IR  photometric  and
spectroscopic observations in the past \citep{mun90, tat03}. From the pulsation period of about 745 d determined by $JHK$-band
photometry, the distance from the Sun has been estimated to be 2.7 kpc using the period-luminosity relation.
 The radius of the photosphere of the mira (M6III; $T_e=2800$ K) is estimated to be $3\times 10^{13}$ cm.
 SiO masers are formed at a few stellar radii of the photosphere, 
 which has well been determined for similar stars in the past from VLBI observations (e.g., \cite{kam10})  
 and SiO maser pumping models  (e. g., \cite{lan84,buj94}). 
 
\citet{mun90} analyzed the B-band light curve 
during the past  75 years and derived a "plausible" orbital period as 43 ($\pm 5$) yr, 
assuming that the  time variation  of the $B$-band magnitude
is due to dust extinction of the light of the white-dwarf orbiting  the mira. 
This period constrains a semi-major axis
of the white-dwarf orbit into a narrow range as $a=14.0$ -- 16.4 AU ($\sim 2.1$ -- $2.4\times 10^{14}$ cm), depending on the white-dwarf mass 
in a range between 0.5 and 1.4 $M_{\odot}$ and an assumed mass of 1 $M_{\odot}$ for the mira (see Table 5 of  \cite{mun90}).  
 The semi-major axis length does not vary much for the more plausible mass of 1.5 -- 2 $M_{\odot}$ for the $P\sim 750$ d mira
(e.g., see equation (5) of \cite{vas93}), 
because it depends on the total mass of the binary system only in the power of $1/3$.

We adopt the length between the white dwarf and the mira of $2\times 10^{14}$ cm at the time of burst.
This length  locates  the  SiO masing region (located at a few stellar radii of the mira) closest to
the white dwarf to be at the halfway between the mira and white dwarf,  i.e., at the point separated both by about $10^{14}$ cm from the mira and from the white dwarf. 
 In this case,  the shock, which propagates with a velocity of 3000 km s$^{-1}$ (suggested from the maximum line width of the H$\alpha$ line; \cite{the10}),
reaches in 4 days at the SiO masing region facing toward the white dwarf. Furthermore, the shock
crosses the whole SiO masing shell in a time scale of about 2 weeks after the burst. However, it is likely that
the shock is gradually decelerated depending on the density of the  ambient medium. 
Molecular gases in the SiO emitting region are also irradiated by UV and soft X-ray radiation.
 We  first present a schematic view of the shock propagation revealed from the present SiO maser observations, 
and then consider it more quantitatively in the following sections.

\subsection{Interpretation of the SiO line profile variation \label{sec:MODEL}}
According to the standard theory of classic novae \citep{hac05,hac06},
nova outburst first starts from the surface of the white dwarf, and creates optically thick "dynamical photosphere"
at a radius of about  $10^{13}$ cm, when the optical light curve reaches the maximum.
Then the hot stellar wind  develops behind the shock front, and 
the photosphere gradually shrinks in parallel with declining the optical light curve (with $t^{-1.75}$). 
In the later phase ($\sim 40$ d for V407 Cyg), the light curve declines more rapidly (with $t^{-3.5}$), 
when thermonuclear reactions cease on the surface of the white dwarf.
It is natural that  UV radiation from the shock dissociates the majority of molecules in the optically thin region
around the mira, except molecules in the shadow of the red giant.
Taking into account the UV optical depth due to dust grain (at 1000 A; $\tau_{\rm 1000}= 4.6 \times 10^{-21} (N_{\rm H_2}$/cm$^{-2}$),
where $N_{\rm H_2}$ is the column density of hydrogen molecules),
we infer that the UV radiation does not penetrate the SiO masing gas with the H$_2$ number density of about $10^9$ cm$^{-3}$
and the scale length of more than $10^{12}$ cm.  Therefore, we
can safely assume that most SiO molecules in the SiO masing region survive from UV radiation of the nova, 
though the real situation may strongly depends on the amount of dust grains
in the SiO masing region and on the clumpiness of the outflowing gas.

A major issue for making a model is how to interpret the split of the SiO line emission
into the red- and blue-shifted sides of the systemic velocity.
In  this paper, we consider that the origin of this velocity split of about 10 km s$^{-1}$
is  due to  the rotational motion of the mira envelope synchronized with the binary orbital motion;  
the red- and blue-shift emissions come from the receding 
and approaching sides of the thick "rotating disk", which is seen in edge-on view from observer (upper panel of Figure 3).  
In this model, the position of the white dwarf must be located at the approaching side 
of the rotational disk to the observer (see Figure 3), 
because the blue-shifted emission disappeared soon after the burst.

Upper panel of figure 3 illustrates the location of the shock front 12 days after the nova outburst.
At this moment, SiO masing in the vicinity of the white dwarf is all  terminated by the shock.
The physical conditions required to maser pumping are
broken due to the shock first at the near side to the white dwarf.  Therefore, at this moment, 
the SiO maser emission was left only at the opposite side to the white dwarf, 
which is a receding part (a red-shifted part) of the rotating disk in this model (indicated by the arrow).  
The shock terminated all the masing gas $\sim 25$ days later. 

\begin{figure}
\begin{center}
\FigureFile(60mm,70mm){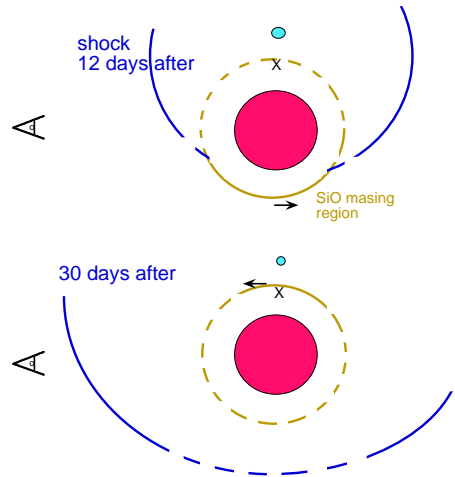}
\end{center}
\caption{ A schematic view explaining the variation of the SiO maser line profiles. Upper panel shows the
location of the shock front  (blue curve) 12 days after the nova explosion, and lower panel 
that of the shock front 30 days after. The red circle indicates the photosphere of the mira,
and the small  light-blue circle indicates the white dwarf (they constitutes a detached binary system).
The orbital plane of the white dwarf  is on this sheet, and
 the observer is located at the infinitely distant point on the left .
The brown circle indicates the SiO masing region, and teared parts indicate the termination 
of masing due to the shock.  The shock due to nova outburst 
terminates masing first at the near side of a masing shell,  
so that the SiO maser emission was left only at the receding part (a red-shifted part) of the rotating disk 
12 days after (indicated by the arrow).  The shock destroys all the masing gas $\sim 25$ days after. 
However, the masing gas is replenished 30 days after by the out flow from the red giant 
through the first Lagrange point (L$_1$; shown as a cross), and makes the blue-shifted component
due to binary rotation.  The pulsation amplitude of the mira is inferred to be about 20 \% of its radius\citep{rei02}. 
\label{fig:3}
}
\end{figure}

The present observation revealed that the blue-shifted components are quickly recovered after the burst.
If we look the observed line profiles carefully,  several  weak  peaks 
at $V_{lsr}=-32$--$ -27$ km s$^{-1}$ are recognizable in the line profile after March 22.
Therefore, the replenishing seems to have started already on 12th days after the outburst,
though once the blue-shifted emission in the observed spectrum almost disappeared on April 6.
If we assume that the velocity of the gas outflow from the mira is 20 km s$^{-1}$,
the gas passes through the  length as small as $5\times  10^{12}$ cm in 30 days. 
It must be a thickness of the maser shell, and
the real coherence length of the masing region (on the line of sight) must be several times this value.
This model fits well to the current observational view of SiO masers that the maser amplification occurs tangentially 
to the radial outflow \citep{dia94}.

The reason of the quick recovery of the blue-shifted emission  is that it is
related to the mass outflow through the first Lagrange point  (L$_1$) of the binary.
A number of SPH numerical simulations were performed for the gas flows of symbiotic binaries
 (for example, see \cite{moh07}). According to  their calculations, the outflowing gas shapes 
 in  expanding spirals originating from L$_1$, at which the gravity 
 balances with the centrifugal force.  It is relatively easier for the radiation/dust driven outflow to escape from the 
 gravity of the red giant through the first Lagrange point due to its low potential barrier. 
 It is most  likely that the replenishing with the molecular gas in the circumstellar envelope starts from the region 
 near L$_1$ after the nova outburst. 
 A schematic view presented in the present paper can offer a reasonable explanation of the 
profile variation of SiO masers after the nova burst in V407 Cyg. 



\subsection{Penetration of the nova shock into the mira envelope}
Observations of the  $\gamma$-ray emission concurrent with the V407 Cyg outburst
offer another useful information on the interaction between nova shock and slow wind of the mira.
The Fermi collaboration \citep{the10} explained that $\gamma$-ray emission, which has been detected for about first two weeks after the nova outburst,
was produced  in the shock by two mechanisms: $\pi^{0}$ decay as a result of the $p$-$p$ interaction,
or  inverse Compton scattering of the mira infrared radiation. Both requires high energy ($\sim$ GeV)
particles which were accelerated by a sudden increase of  magnetic field strength due to the interaction between the nova shell with a velocity of 3000 km s$^{-1}$ 
and ambient medium of the mira wind. Here the ambient medium has a H$_2$ number density of about $5\times 10^7$ cm$^{-3}$
at a distance of $1.0\times 10^{14}$ cm from the mira 
with a mass loss rate of $3\times 10^{-7}\ M_{\odot}$  y$^{-1}$.
The mass of the ejected material is deduced to be approximately $10^{-6}\ M_{\odot}$ 
(for the O/Ne nova with a white dwarf mass of 1.2 $M_{\odot}$; \cite{sta00}).

A spherical shock with a velocity of 3000 km s$^{-1}$ will gradually be decelerated 
by sweeping the wind of the mira.  The velocity decreases to a half  
if the shock sweeps approximately the same mass as the ejecta. 
The deceleration scale length at a point in the envelope can be approximated by
$  l_{dec} \approx   v_{\rm wind} \times \  (M_{\rm ej}/ \ \dot M_{\rm mira} \ ) \times  (R/D)^2 $
for a spherically symmetric mira wind, where $v_{\rm wind}$, $M_{\rm ej}$, and $\ \dot M_{mira}\ $ are  the velocity of the mira wind,
the mass of nova ejecta, the mass loss rate of the mira, respectively, and $R$ and $D$ are distances from the center of the mira and from the center of 
the white dwarf, respectively. This equation gives  
the  deceleration scale  length to be $10^{14}$ cm for the hydrogen number density 
of the ambient medium of $N_{H_2}= 5\times 10^7$ cm$^{-3}$ at $R=D=10^{14}$ cm  
(for  $M \ _{\rm ej}=1.0\times  10^{-6} M_{\odot}$ y$^{-1}$,  $ \ \dot M \ _{\rm mira}=3\times 10^{-7} M_{\odot}$ y$^{-1}$,
and $v_{\rm wind}= 10$  km s$^{-1}$). Therefore, the  shock can
sustain its speed of 3000 km s$^{-1}$ at the halfway between the white dwarf and the mira. 
 The deceleration length scale decreases rapidly to $10^{13}$ cm  if it hits
the gas of  $N_{H_2}= 2\times 10^8$ cm$^{-3}$ at $R= 5\times 10^{13}$ cm and $D=1.5\times 10^{14}$ cm. 
In fact, observed variations of the H$_{\alpha}$ wing  (see figure 5 of  \cite{mun10}) indicate that the shock sustained 
the velocity above 2000  km s$^{-1}$ during the first 5 days, and it was decelerated to the velocity of about 1000 km s$^{-1}$ 17 days after.
Therefore the shock front propagated in the distance of  approximately  $3 \times 10^{14}$ cm (on the line of sight)  
in the first 3 weeks, which is roughly consistent with the estimation of the shock deceleration length scale present above.

In reality,  the gas in the mira wind is more or less clumpy (e.g., \cite{gra97}).
The shock can blow out the  SiO masing clumps with $N_{H_2}=10^9$ cm$^{-3}$
with a diameter of $10^{13}$ cm at the half way between the mira and the white dwarf  ($R\sim 10^{14}$ cm).
Furthermore, the gas density would decrease more rapidly toward the polar axis
of the binary orbit  than to equatorial directions,  as calculated by \citet{moh07}.  
Because shock deceleration is less severe toward the polar directions, 
the shock expands bipolarly. 
Such a bipolar expansion of the nova shock
has been observed by VLBI  for the recurrent nova, RS Oph \citep{obr06}, 
for which the energy of the event is comparable with that of the V407 Cyg outburst \citep{sok06}.


Strong X-ray emission from V407 Cyg was also observed  two weeks after the burst,
and it continued more than a month \citep{nel10}. This emission is inferred to come from the 
shock which developed  toward the polar axis initially, and wrapped over the inner dense envelope of the mira later.
From the XRT  photon counts of the Swift X-ray telescope (see Figure 1 of \cite{the10}),
we estimated the soft X-ray flux  passing the unit area at $R=10^{14}$ cm
to be $1.8 \times 10^{13}$ photon cm$^{-2}$ s$^{-1}$. 
Because the optical depth of the soft X ray ($<0.6$ keV) is about 1/20 of 
$\tau$(dust) at 1000 A (e.g., see \cite{deg90}), it can penetrate into the SiO masing gas.
Soft-X-ray photons ionize H$_2$ molecules, producing  mostly H$_2^+$, which are transformed in other form
of molecular ions (H$_3^+$ and H$_3$O$^+$), and quickly recombine with electrons.
Since soft-X-ray ionization of He is approximately 10 times efficient than the ionization of H$_2$ \citep{gla73},
 the most efficient source of electron is He  in the molecular environment under soft-X ray irradiation.  
Because the recombination of He$^+$ with electron and the reaction of He$^+$  with H$_2$ are both slow, 
the most efficient route to return He$^+$ to He is the reaction with H$_2$O
(He$^+$ + H$_2$O $\to$ OH$^+$ + H + He) with a reaction rate of $3 \times 10^{-10}$ s$^{-1}$. 
Therefore  electron density in the gas can be estimated from the He ionization, and the ion-molecular reaction with H$_2$O.
We estimated that  the above amount of  
soft X-ray irradiation can produce an electron density of  $3\times 10^{4}$ cm$^{-3}$ 
in the SiO masing clumpy shell with a $5\times 10^{12}$ cm thickness
in a time scale of $10^4$ sec.  It also heats the gas clump by a few hundred degree per day, but this 
does not induce severe dynamical change  of SiO masing gas of the hydrogen density of $10^9$ cm$^{-3}$.
However, the molecular excitation is significantly influenced by the increase of the electron density,
which may terminate SiO masers in an assumption that electron collision  thermalizes the populations of maser levels.

These estimations suggest that the nova shock wraps outside of the entire SiO masing region (of about $r=10^{14}$ cm)  in two weeks,
 and the soft X-ray penetrates into the  SiO masing clumps at the opposite side of the mira to the white dwarf
 and  severely changes the molecular excitations there. 
It is likely that SiO masers of V407 Cyg at the higher velocity side (i.e., at the opposite side of the white dwarf) were
vanished in two weeks after the nova outburst by soft X-ray irradiation. 
The coincidence between the time of soft X-ray emergence and the time of SiO maser disappearance, 
seems to support this idea.


    
\section{Summary}
SiO maser emission from the symbiotic star, V407 Cyg, was monitored for 2.5 months 
after the nova outburst occurred on 2010 March 10. 
The observations showed that  the $J=1$--0 $v=2$ line gradually disappeared in two weeks, and
a new  component, which was shifted in velocity by about 5 km s$^{-1}$  from the disappeared component, 
emerged three weeks after the burst. The $J=1$--0 $v=1$
line was marginally detected during this period. These observations is reconcilable with the
 shock propagation model for classical novae. The shock stroked through the SiO masing region and 
 terminated masers. However, the masing gas was replenished quickly in three weeks after the nova outburst.
 The recovered area is inferred to appear near the first Lagrange point, through which  matters ejected from the mira 
can move with relatively low kinematic energy. 
 
The picture presented here is roughly consistent with the shock propagation model 
which was proposed for the $\gamma$ ray emission concurrently observed with V407 outburst.
Since there still remain many free parameters for mass-outflowing binaries, 
high spatial-resolution observations such as very long baseline interferometry (VLBI) 
for the continuum and SiO masers will be essential to constrain the models.
Hopefully the VLBI observations give us the detailed geometry of the system composing of the white dwarf, the AGB star,
and the mass outflow which is manifested by SiO masers. 

\

This research was made use of the  NRO project time \# 493004 for the Nobeyama 45m telescope. 
This research was partially supported by a Grant-in-Aid for Scientific Research from
Japan Society for the Promotion of Sciences (20540234), and by a grant awarded to JN
from the Research Grants Council of Hong Kong (HKU  704710P; 703308P; 702208P). 

\end{document}